\documentclass[runningheads]{llncs}
\usepackage{savesym}
\savesymbol{EUR}
\usepackage{eurosym}
\usepackage{booktabs}
\usepackage{color}
\usepackage{graphicx}
\PassOptionsToPackage{hyphens}{url}
\usepackage{hyperref}
\usepackage[utf8]{inputenc}
\usepackage[numbers,sort&compress]{natbib}
\usepackage{siunitx}
\usepackage{subfig}
\restoresymbol{marvosym}{EUR}

\usepackage[show]{chato-notes}
\graphicspath{ {imgs/} }

\begin{document}
\title{Facebook Ads: Politics of Migration in Italy}
\author{Arthur Capozzi\thanks{After first author, the author names are in alphabetical order.} \and
Gianmarco De Francisci Morales \and
Yelena Mejova \and\\
Corrado Monti \and
Andr\'{e} Panisson \and
Daniela Paolotti}
 
\authorrunning{A. Capozzi et al.}
\institute{ISI Foundation, Turin, Italy, 
%\email{arthur.capozzi@gmail.com}\\
\url{isi.it}}
\maketitle              % typeset the header of the contribution

% Author order
% Arthur Capozzi, Gianmarco De Francisci Morales, Yelena Mejova, Corrado Monti, Andre Panisson, Daniela Paolotti

\begin{abstract}
%The abstract should briefly summarize the contents of the paper in 15--250 words.

Targeted online advertising is on the forefront of political communication, allowing hyper-local advertising campaigns around elections and issues.
In this study, we employ a new resource for political ad monitoring -- Facebook Ads Library -- to examine advertising concerning the issue of immigration in Italy.
A crucial topic in Italian politics, it has recently been a focus of several populist movements, some of which have adopted social media as a powerful tool for voter engagement.
Indeed, we find evidence of targeting by the parties both in terms of geography and demographics (age and gender).
For instance, Five Star Movement reaches a younger audience when advertising about immigration, while other parties' ads have a more male audience when advertising on this issue.
We also notice a marked rise in advertising volume around elections, as well as a shift to more general audience.
Thus, we illustrate political advertising targeting that likely has an impact on public opinion on a topic involving potentially vulnerable populations, and urge the research community to include online advertising in the monitoring of public discourse.

\end{abstract}

\section{Introduction}

According to the International Organization for Migration, almost 130k migrants have arrived in Europe in 2019.\footnote{\url{https://migration.iom.int/europe?type=arrivals}}
Global migration is a systemic challenge for Europe~\cite{joppke1998challenge}, and for Italy in particular being on the Mediterranean route~\cite{frontex2020migratory}.
%Migration is influenced by a combination of economic, environmental, political, and social factors, and in turn generates changes in each of those areas.
%These major changes require the involvement of governments to manage the migratory phoenomenon~\citep{boghean2016phenomenon}.
Therefore, it is not surprising that migration is a central issue in European and Italian politics~\cite{human2020events}.

At the same time, Europe and the world in general have seen a resurgence of nationalism with a populist derive, such as what seen in USA, Brazil, Philippines, Turkey, UK, Hungary, and Italy to name a few~\citep{smith2013nations}.
Nationalist parties often espouse nativist positions, denouncing the negative effects of migration, and emphasizing the loss of cultural identity~\citep{falk2010invasion}.
Indeed, migration is a controversial issue often used as a political tool by parties across the aisle.
In this work, we focus on the Italian debate around migration, which has been of paramount importance in recent history.\footnote{\url{https://www.bbc.com/news/world-europe-43167699}\\ \url{https://www.politico.eu/article/italy-immigration-debate-facts-dont-matter}}

Some of the parties have risen to success by embracing the new communication technologies available on the Web and social media~\citep{beer2019landscape}.
The Five Star Movement (\emph{Movimento 5 Stelle}, M5S)~\citep{natale2014web} is a notorious example of this trend, with its focus on anti-establishment rhetoric paired with an organizational focus around new communication technologies.
In contrast with traditional media which focus on mass communication (TV and news), social media allows direct and personal communication.
This \emph{micro-targeting} feature is highly controversial and has caused wide backlash towards Facebook~\cite{hern2019facebook}.

Given this socio-political context, our main research aim is to study the \emph{political messaging around migration in Italy via Facebook ads}.
In particular, we look for evidence of micro-targeting and analyze different focus of the major parties on Italian stage.
We find that different parties have different demographic foci, and the target audience shifts during events such as elections.
In addition, we find evidence that nationalist parties focus on more male audience for their migration ads compared to their normal targets.
Finally, we discuss the advantages and limitations of this methodology, and future integration of online data in political discourse analysis.

\section{Background}  % Related Work
\label{sec:background}
% Facebook Ads and other ad targeting studies
\paragraph{Political advertising.}
As efforts toward transparency have recently resulted in Facebook and other major platforms releasing information about political advertising, the research is just beginning to measure the extent and impact of online targeting on vulnerable groups~\cite{beer2019landscape}.
For instance, Hegelich \& Serrano \cite{hegelich2019microtargeting} examined political ads made available through Facebook Ad Library API and the Google Cloud BigQuery API, finding targeting differences among major German political parties.
While some parties published ads with broad appeal, others had specific targets, such as women of working age.
However, no differentiation between topical focus was made in this analysis.
Other sources of ad-related datasets become available as political events unfold, such as in the case of Ribeiro et al.~\cite{ribeiro2019microtargeting} who examined a dataset of 3,517 Facebook ads that were linked to a Russian group Internet Research Agency (IRA).
Targeted to U.S. audience, the ads achieved a higher click through rate (CTR) than usual Facebook ads.
After running a survey on the most impactful ads, the authors found that many of these ads were severely divisive, and ``generated strongly varied opinions across the two ideological groups of liberals and conservatives''.
In particular, these ads were targeted to those who were more likely to believe, and approve and subsequently less likely to report or identify false claims in them.
However, the Facebook Ads Library API has been shown to be unstable during the 2019 European Parliament Election \cite{mozilla2019data,rosenberg2019ad} -- this study is the latest attempt to negotiate the API to extract useful information about political advertising and targeting.

% Story of Italy and its political situation
\paragraph{Italian parties and migration.}
% Understanding the advertising around the issue of immigration is especially important in today's political environment.
% In 2009, Van der Brug \& Van Spanje \cite{van2009immigration} surveyed immigration policies of 82 political parties in 14 European countries, and concluded that ``large groups of citizens are not represented by any parties, in particular those who are left-wing on socio-economic issues and right-wing on cultural issues''.
% Since then, Italy has seen an upsurge in populist politicians~\cite{van2015populist}.

In the time period we analyzed, March 2019 to March 2020, the main Italian parties are the Five-Star Movement (\emph{Movimento 5 Stelle}, M5S), the League (Lega), the Democratic Party (\emph{Partito Democratico}, PD).
M5S was the largest party in Italy at the previous national elections in March 2018; it currently holds about one third of the parliament, while Lega, PD, and Forza Italia hold about one fifth each.
M5S has been unanimously described as anti-establishment and populist; however, its position on the left-right spectrum has been harder to describe~\cite{mosca2019beyond}.
\citet{emanuele2020times} found that ``M5S voters are leftist on the economy but quite close to right-wing voters on Europe and immigration''.
M5S's positions on migration is wavering~\cite{pasini20196}, mixing humanitarianism with defense of national borders~\cite{mosca2019beyond} -- for instance, they accused NGOs of increasing illegal immigration by rescuing migrants at sea~\cite{coticchia2020populist}.
% While M5S has never been appealing to etno-cultural, identitarian conceptions, Ivaldi et al.~\cite{ivaldi2017varieties} noted that ``key personalities of the party have used xenophobic language against migrants'' and, 
According to~\citet{gerbaudo2017reclaiming} this strategy has the goal of competing on the issue with Lega.

Lega under Salvini is in fact keeping a strong anti-immigration and nativist focus, adopting ``stop the invasion'' as a slogan, calling for immediate  repatriations, and depicting Islam as a threat to Italian Christian identity~\cite{ivaldi2017varieties}.
As the interior minister in June 2018, Salvini declared Italian ports closed to NGO ships rescuing migrants~\cite{cusumano2020deep}.
Brothers of Italy (\emph{Fratelli d'Italia}, FdI) holds similar positions, and was allied with Lega and Forza Italia in the center right coalition in 2018. % tend to use less charged tones, and display a more ``pragmatist'' views; however, they often
%  exploited migration against their opponents.~\cite{gianfreda2018politicization}

The Democratic Party, who has dominated Italian government coalitions after 2013, is the main target of such critiques from other parties.
PD adopted a strategy of ``setting up a decentralized system for the management of asylum requests''~\cite{diamond2019italian}. In their view, migration must be managed more than stopped: for instance, they aimed at reducing migrant flows to Italy~\cite{guardian2017}, through agreements with Libya~\cite{bulli2018immigration} and by negotiating the redistribution of migrants with the rest of EU. However, they kept a mostly humanitarian position in public discourse about  rescuing operations at sea. 
During the analyzed time period, former prime minister Matteo Renzi split from PD to form a new formation, Italia Viva (IV).

\section{Data}

The data for this collection comes from Facebook Ads Library, which provides an API\footnote{\url{https://www.facebook.com/ads/library/api}} for accessing advertisement shown on Facebook and Instagram platforms about ``social issues, elections or politics'' since March 2019.
On March~30, 2020 we run a query to collect all advertisements originating from Italy containing keywords related to migrants, immigrants, and refugees (see Appendix~\ref{sec:appendix} for the exact keywords).
We use keywords in Italian language because we are interested in the domestic perception of the topic of migration.
The keywords were compiled using an iterative approach, by starting from ``migrante'' (migrant) and ``immigrato'' (immigrant), and then adding keywords to the set using FastText embedding similarity~\cite{grave2018learning}.
In particular, we search for ads appearing only on Facebook (not Instagram).

The API returns ads that were published since March 2019, including those which are running at the time of the collection, and excluding those which have been deleted (mostly removed by the platform).
For each ad, the API provides the ad ID, title, and body, and URL, its creation time and the time span of the campaign, the Facebook page authoring the ad, as well as a funding entity.
Additional information includes a range for the cost (as well as its currency) and impressions of the ad.
The API also provides a distribution over impressions broken down by gender (male, female, unknown), age (7 groups), and location (down to region in Italy).
We restrict our analysis by considering only impressions localized in Italy.
In further analysis of age, cost and impression information, which come in a range, we take the average of the end points of the range, and for open-ended ranges such as 65+ we take the known closed end point (in this case, 65).
Note that in further analysis we focus on the expenditure and impressions, rather than the raw number of ads, as the same message can be presented in several ads, thus making a single ad not a meaningful unit of analysis.
For example, to compute the cost per thousand impressions for an author, we divide their total expenditure on all their campaigns by the total number of impressions received.
Upon manual examination of top authors, we exclude 7 who are not relevant for the aim of our analysis, such as brands and emigration advertising.\footnote{The full list of authors excluded is reported in Appendix~\ref{sec:appendix}}
The dataset consists of \num{2312} ads from \num{733} unique pages.

% May want to describe the "background" collection, if we use it

To better understand the pages in our dataset, we query WikiData\footnote{\url{https://www.wikidata.org}} with all n-grams of words in the name of each page.
We restricted the matched entities in WikiData to a specific set of categories: parties, politicians, journalists, and NGOs.
In all the cases where we had more than one match, we manually disambiguate the matched entities and select the relevant one.
To further refine our data, we match the remaining pages against a list of local Italian politicians.\footnote{\url{https://dait.interno.gov.it/elezioni/open-data/dati-amministratori-locali-carica-al-31-dicembre-2018}}
The process results in 249 pages identified as individual politicians and 53 pages identified as parties (including regional branches).

We then consider the pages of politicians affiliated with one of the four parties in Italy most present in our data -- Democratic Party (PD), League (Lega), 5 Stars Movement (M5S), Brothers of Italy (Fratelli d'Italia, FdI), and Italy Alive (Italia Viva, IV) -- resulting in 208 pages.
For each page, we retrieve all ads without keyword constraint, \num{17014} ads in total, which we later use as a general political baseline.

\section{Results}

\subsection{Characterizing Advertising around Migration}

%Who are the pages getting most impressions, who are very correlated with top spenders they get, calculate correlation. Gini coefficient, how skewed it is.

\begin{figure}[t]
    \centering
    \includegraphics[width=0.8\textwidth]{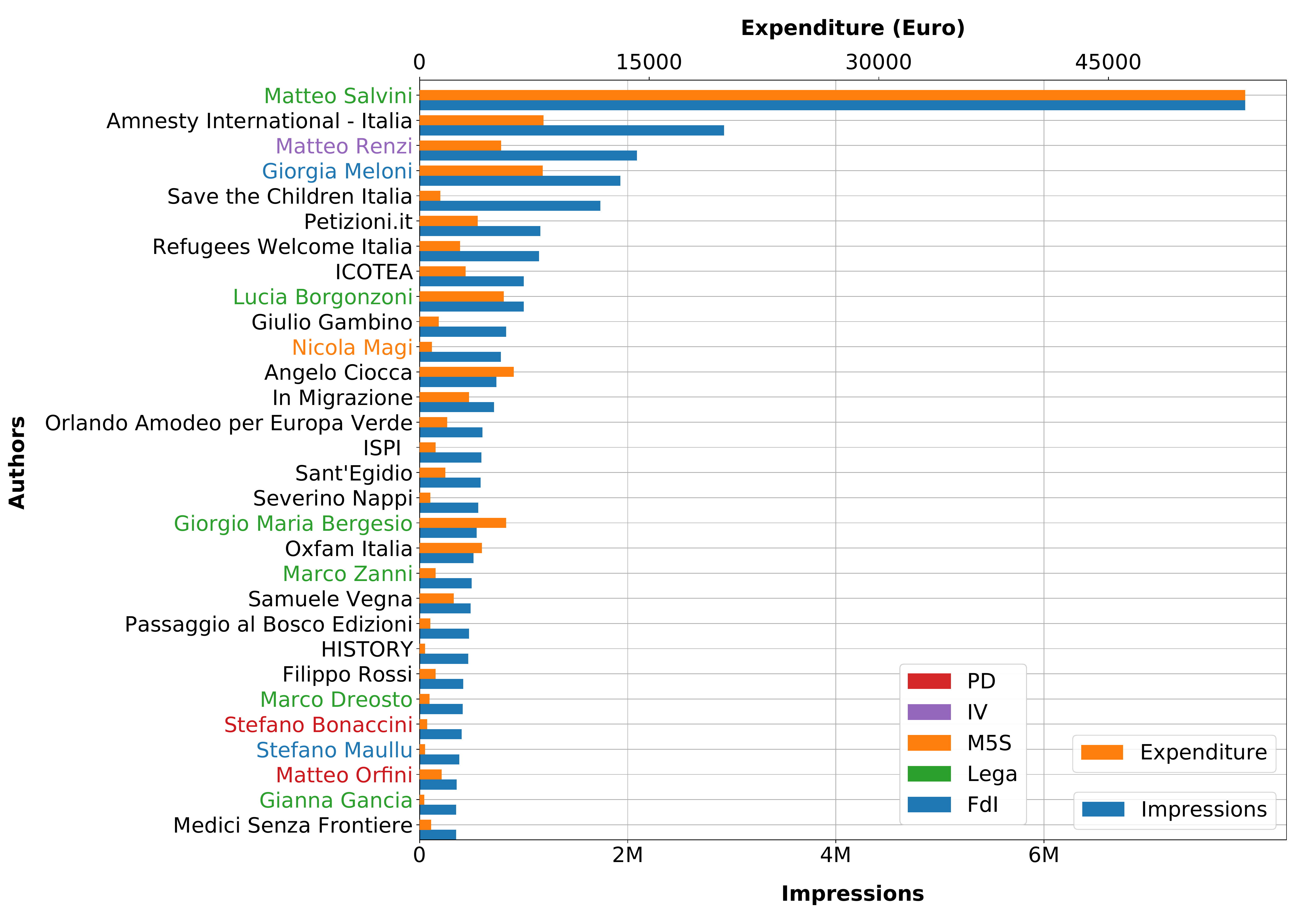}
    \caption{Authors by impressions and expenditure per page (top 30 for impressions).} \label{fig:hbar_plot_average_impressions_spent_per_page}
    \vspace{-\baselineskip}
\end{figure}

We begin by considering the top spenders in our data, alongside those whose ads received the most impressions.
These statistics are plotted in Figure~\ref{fig:hbar_plot_average_impressions_spent_per_page}, with bottom x axis showing the impressions and top the spending, summed over all ads by the given author.
The most prolific author is Matteo Salvini, the leader of the Lega party -- a major right-wing nationalist party -- having spent in total over \num{50000} Euros and attracted almost \num{8000000} impressions.
Interestingly, we find other party leaders in the top places: the third most prolific author is Matteo Renzi (former leader of PD, current leader of IV), ahead of Giorgia Meloni (leader of FdI), despite spending less, pointing to a different cost per impression.
The second most prolific author is an NGO, the Italian branch of Amnesty International, followed by others such as Save the Children Italia and Refugees Welcome Italia.
As we detailed in Section~\ref{sec:background}, NGOs have been at the center of political controversies for their rescuing operations.
Other authors include a journalist (Giulio Gambino), a petition website (petizioni.it), and an online learning website (ICOTEA).

Thus, we find that the biggest spender on ads related to migration comes from the political domain.
In particular, Matteo Salvini being the largest spender on ads about immigration is not surprising: first, his party has been recognized as ``one of the first political entrepreneurs of anti-immigration sentiments in the Italian arena''~\cite{bulli2018immigration}; moreover, political commenters recognized Matteo Salvini as successful on leveraging social media to promote his views.%
\footnote{See for instance \href{https://www.theatlantic.com/international/archive/2019/09/matteo-salvini-italy-populist-playbook/597298/}{The Atlantic, \emph{``The New Populist Playbook''}} and \href{https://www.reuters.com/article/us-italy-politics-salvini-socialmedia/chestnuts-swagger-and-good-grammar-how-italys-captain-builds-his-brand-idUSKCN1MS1S6}{Reuters.com, \emph{``Chestnuts, swagger and good grammar: how Italy's 'Captain' builds his brand''}}}
% \mynote{YM: CORRADO: Write about Salvini and Lega and their focus on migration as a ideological leverage, cite a major newspaper\footnote{\url{https://www.leganord.org/immigrazione-linee-guida}}}
Facebook has been recognized in previous analysis~\cite{mazzoleni2018socially} as an important tool also for the other party leaders we find in the top positions.

The resources used for this political advertising make for an unequal distribution of impressions.
The Gini coefficient of impressions per page is \num{0.465}, while for individual ads even larger at \num{0.800}, thus indicating a highly skewed distribution of impressions across the field.

% \mynote{YM: Add Gini coefficient of impressions}.
% \mynote{Arthur: per page = 0.4651924648584149. Per ads = 0.800294693022333}.
% The biggest ad: https://www.facebook.com/ads/library/?id=303357167226217

Considering the efficiency of the campaigns, we find that most are able to achieve better impression per euro spent ratios than the top spender.
Whereas Matteo Salvini paid on average \num{6.8} ([\num{3.2}, \num{11}]) Euros per thousand impressions (CPM), everybody else in the top 30 has spent \num{4.1} ([\num{0.6}, \num{10.7}]) Euros per thousand impressions.
The most expensive impressions were purchased by Giorgio Maria Bergesio (local politician from Lega party) at \num{10.3} CPM, while the cheapest by Save the Children Italia at \num{0.8} CPM.
We address whether this difference may be linked to microtargeting in the next section.

% Range: [ min exp / max impr --- max exp / min impr ]

%exp       impr
%100-1000  50-100
%100-1000  100-500
%100-1000  50-100
%(550*3)/(75 + 150 + 75) -- avg cpi
%(100*3)/(100+500+100) -- min cpi
%(1000*3)/(50+100+50) -- max cpi

% SALVINI
% MIN:  0.0032104267215448015
% AVG:  0.006798782555167045
% MAX:  0.011041534864530326

% ALL but SALVINI
% MIN:  0.0005977192808109462
% AVG:  0.004133431157706486
% MAX:  0.010739413339018867

% Considering top 30 authors for Impressions:
% AUTHOR
% MIN. min:  ("Sant'Egidio", 0.00013477670196019236)
% MIN. max:  ('Lucia Borgonzoni', 0.005)

% AVG. min:  ('Save the Children Italia', 0.000775828231169187)
% AVG. max:  ('Giorgio Maria Bergesio', 0.010289671981980095)

% MAX. min:  ('Save the Children Italia', 0.0011498305084745764)
% MAX. max:  ('Giorgio Maria Bergesio', 0.029214477211796246)

%Temporal distribution of the dataset, main events that correspond to spikes.

\begin{figure}[t]
    \centering
    \includegraphics[width=0.85\textwidth]{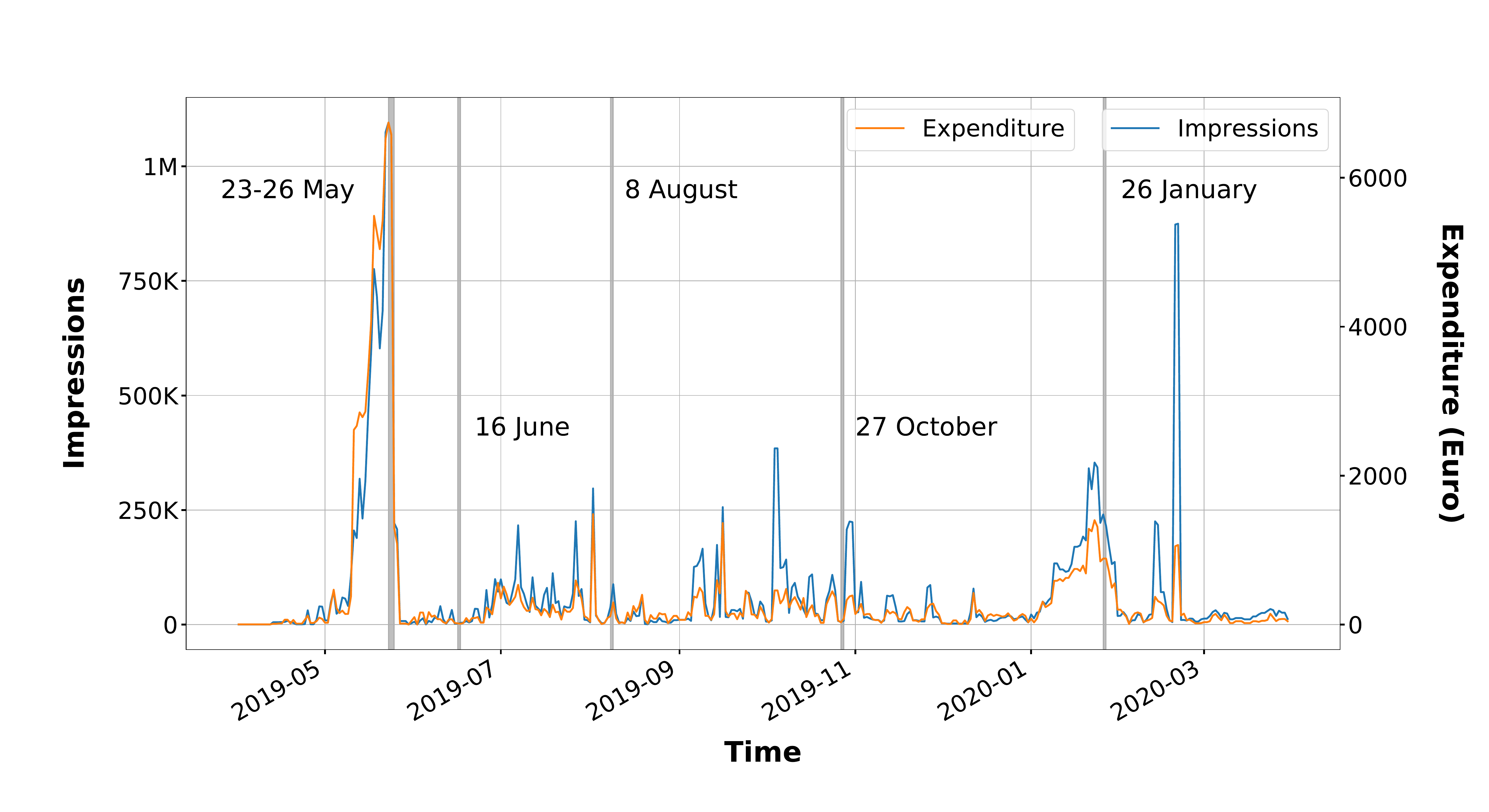}
    \caption{Impressions and expenditure for ads related to migrations, per day. We also highlight important political events with vertical bars: 23-26 May - European and regional (Piedmont) elections; 16 June - Regional elections in Sardinia; 8 Aug - Lega leaves the government; 27 Oct - Regional election in Umbria; 26 Jan - Regional elections in Calabria and Emilia-Romagna. } \label{fig:impressions_and_spend_per_day}
    \vspace{-\baselineskip}
\end{figure}

Figure~\ref{fig:impressions_and_spend_per_day} shows the temporal distribution of impressions and expenditure per day.
Specifically, we divide the total impressions (expenditure) uniformly over the duration of each ad campaign.
The two time series are highly correlated, and are bursty.
We also show major events: all major elections (European elections, as well as regional in Piedmont, Umbria, Calabria, and Emilia-Romagna) and the government crisis.
We observe that some of those events correspond to spikes in ads expenditure.
During these elections, the major players were political parties, which behave differently at different elections, and we explore their behavior in further sections.
The largest spike is in the weeks preceding the European elections on 23-26 May (other regional elections also take place then).

% \mynote{Check out ads on Feb 2020, see what they may have been about.}
% \mynote{The spike is during two days (20-21 Feb.):\\
% '2020-02-19': {'impression': 88889, 'spend': 374},\\
% '2020-02-20': {'impression': 986634, 'spend': 1428},\\
% '2020-02-21': {'impression': 989915, 'spend': 1447},\\
% '2020-02-22': {'impression': 137502, 'spend': 430},\\
% It seems that the spike is due to a couple of Ads created by Matteo Renzi. These ads were active for just two days and they got about 600K impressions ((250K-300K)+(300K-350K)) subdivided in just two days.
% \url{https://www.facebook.com/ads/library/?id=2512680902308111}\\
% \url{https://www.facebook.com/ads/library/?id=630292624437466}}

% 23-26 May:
%   European elections
%   Piedmontese regional election
%   Administrative elections in 57 municipality

% 16 June
%   Administrative election in Sardinia.

% 8 August
%   Lega leaves the government and asks for a return to the polls.

% 27 October
%   Umbrian regional election

% 26 January:
%   Calabria regional election
%   Emilia-Romagna regional election

\begin{figure}[t]
    \centering
    \includegraphics[width=0.85\textwidth]{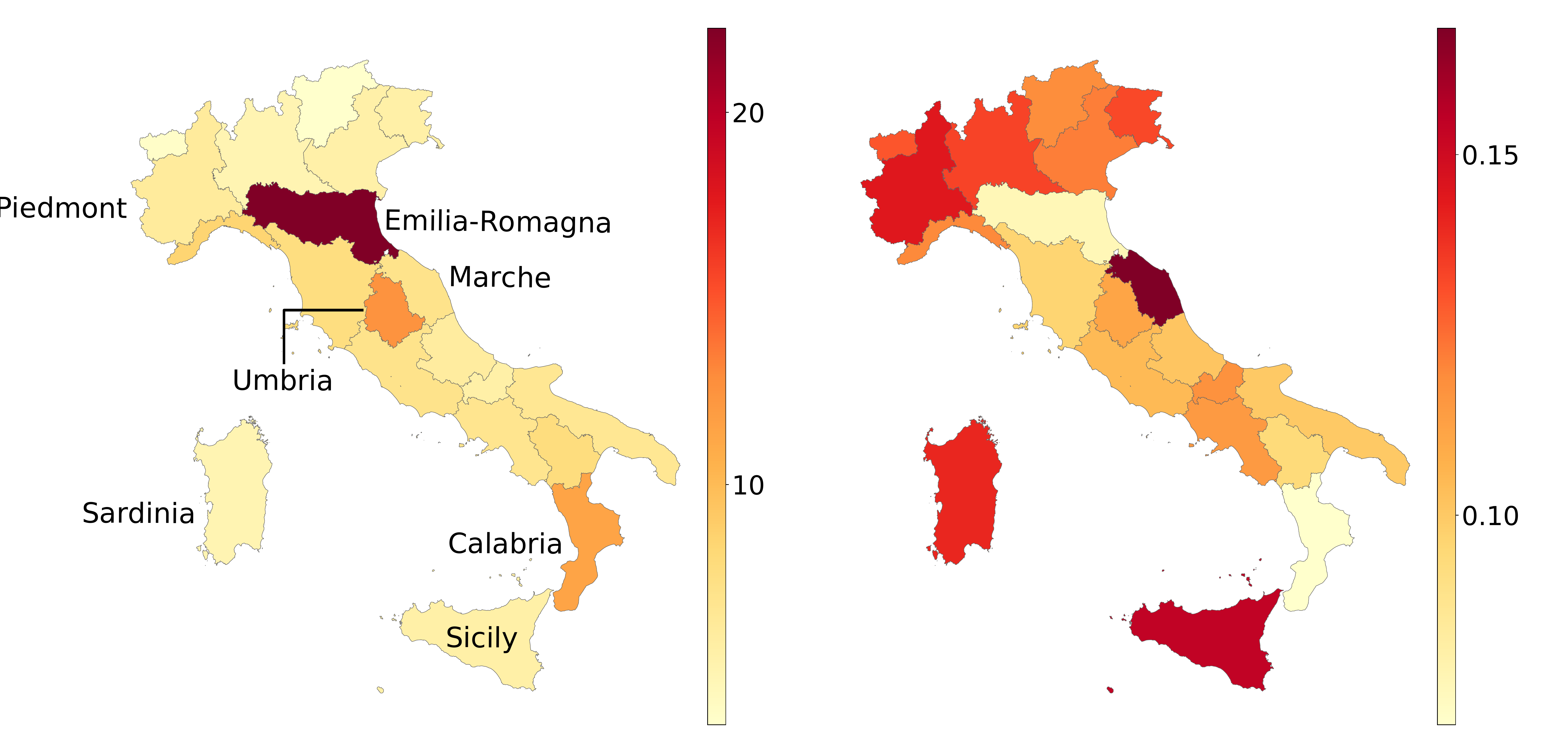}
    \caption{Impressions per capita of all political ads (left), fraction of impressions of migrant-related ads over all political ads (right).} \label{fig:map_impressions_fraction}
    \vspace{-\baselineskip}
\end{figure}

Figure~\ref{fig:map_impressions_fraction} shows the geographical distribution over regions of the number of impressions of political ads normalized by population (left) and of the fraction of impressions of migrant ads (right).
The most impressions per capita of political ads are Emilia-Romagna, (with an average of \num{22.2}), followed by Umbria (\num{12.4}), and Calabria (\num{11.4}).
These are in fact the regions that had regional elections in the reference period.
The fact that these impressions per person are larger than 1 implies that an average user saw several ads in that time.
Emilia-Romagna having the highest number of impressions per capita (despite having a large population) confirms the perceived political importance of their 2020 regional elections:
many observers noted in fact that those local elections were particularly important for Italian politics, as they could affect the national government~\cite{amaro2020salvinis}.
%Note that the normalization by population brings out the smaller regions such as Marche and Umbria. While Umbria had an important local election, we find that a single advertising campaign is responsible for the large portion of the impressions in Marche.
%Geographic distribution. Discuss focus points - the elections, Umbria one has more impressions per user. Foreshadow: Emilia Romagna does not have as much focus on migrants as Umbria.

The map on the right in Figure~\ref{fig:map_impressions_fraction} shows the fraction of political ads that are related to immigration.
We note how there is a geographical gradient in attention to this topic: northern regions and islands (Sardinia and Sicily) show more focus on this topic.
The northern regions are typically viewed as strongholds of Lega,\footnote{\url{https://commons.wikimedia.org/wiki/File:Camera_2018_Partiti.svg}} whose political agenda is strongly focused on immigration.
The prevalence of this topic in Sicily and Sardinia is more surprising.
It is interesting to note that these regions emerged in a survey as having the largest fraction of the population overestimating the presence of immigrants~\cite{saso2019immigrazione}.
Emilia-Romagna and Calabria show the least prevalence of the topic.
% Many observers noted that the topic of immigration was not a major part of the electoral discussion~\citemissing.
In the next section, we find that these regions also display a more generalist targeting: the age and gender distribution of their ads is closer to the population average. %, less focused on older population and males.
Finally, we note that the attention to the topic in Sicily could be connected to the fact that most rescue ships operations happen near its shores.
We also find that a single advertising campaign (Nicola Magi, M5S) is responsible for the large portion of the impressions in Marche (44\%).

%\begin{figure}
%    \includegraphics[width=\textwidth]{figs/geographic_distribution.png}
%    \caption{Geographic distribution of impressions (left), same normalized by total population (right). \inote{Put the names of regions we refer to in the maps.}} \label{fig:map_impressions_and_spend_per_day}
%\end{figure}

\begin{figure}[t]
    \centering
    \includegraphics[width=0.6\textwidth]{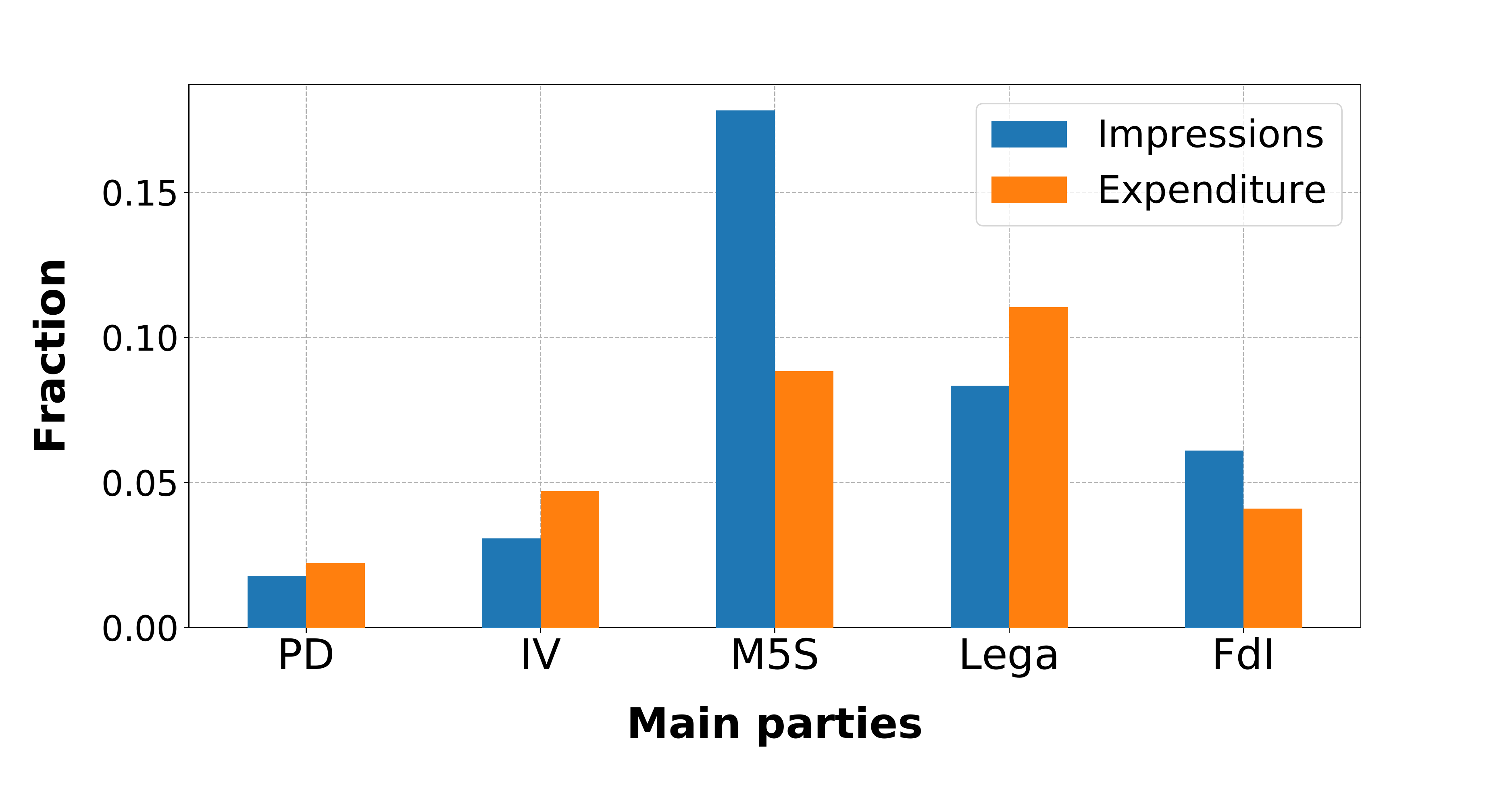}\vspace{-0.3cm}
    \caption{Fraction of advertising budget and reach that each party dedicates to migrant-related topics.} \label{fig:fraction_expenditure_impressions}
    \vspace{-\baselineskip}
\end{figure}

In Figure \ref{fig:fraction_expenditure_impressions} we put into context the expenditure committed and impressions gained by the major political parties, as a proportion of the total political advertising.
We find that Lega and M5S have invested the most into migrant-related ads as a proportion of their total budget.
Among the parties, M5S attains the largest proportion of their viewership from the migrant-related ad impressions.
The party having the least focus on the migrant topic is PD, both in terms of relative budget spent and impressions.
Out of these, the one that has spent the most on migrant-related advertising is Lega spending in total \num{87314} \euro, followed by PD (\num{13786} \euro), FdI (\num{10963} \euro), IV (\num{6023} \euro) and M5S (\num{3533} \euro).

%Topic modeling and content analysis, embeddings?
%Table: top words in topics
%FIND WHERE THE TOPICS AND EMBEDDINGS ARE IN NOTES

Finally, we look at the messages conveyed by each party, by analyzing the text of each ad.
Table~\ref{tab:top_words_per_party} in Appendix~\ref{sec:appendix} shows the top terms used by each party about migration, sorted by odds ratio of frequency in migrant-related ads to all political ads.
First, we find a difference in the way each party speaks about migration: Lega and FdI both focus on ``clandestines'' (i.e., illegal immigrants) in their ads.
Second, all parties speak about themselves, but interestingly Lega and Salvini are often mentioned by their political opponents.
Third, FdI, refers to law, territory, nation, and citizenship, thus emphasizing the defense of Italian sovereignty.
Both FdI and Lega focus on rescue ships operations, through the words ``NGO'' for the former and ``ports'' for the latter\footnote{{E.g., using the hashtag \small \texttt{\#portichiusi}} -- ``close the ports''.}.
% \todo{Corrado: remind as we said before that NGOs became a political target.}
Instead, PD is mentioning words related to rights and duties%
\footnote{The definition of ``rights and duties'' for immigrants is PD's first point of intervention for their integration (see \url{https://www.partitodemocratico.it/archivio/il-pd-e-limmigrazione/}).} %
in relation to immigrants, such as ``law'' and ``rights''.
IV is mostly focusing on ``hate'' (denouncing Salvini's campaign as ``hateful'')\footnote{E.g., \url{https://www.facebook.com/ads/library/?id=2095571910748061}.} as well as connecting immigration to ``security''.

\subsection{Audience Targeting}

\begin{figure}[t]
    \centering
    \includegraphics[width=0.6\textwidth]{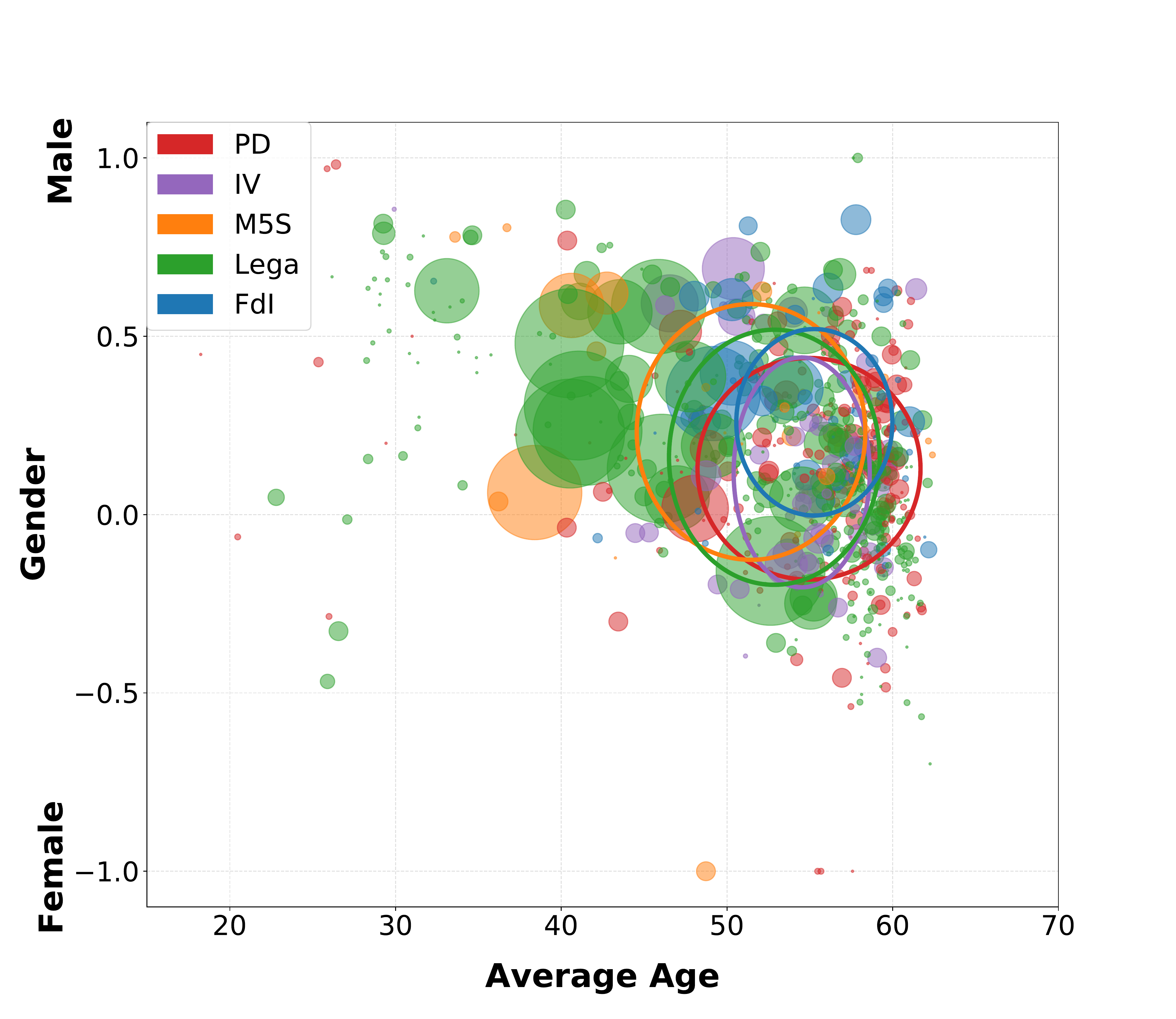}
    \caption{Average reach of individual ads authored by political parties over age and gender, size proportional to impressions. Ellipses represent one standard deviation for each axis around the average for each party (weighted by impressions).
    }
    \label{fig:fireworks}
    \vspace{-\baselineskip}
\end{figure}

\begin{figure}[t]
    \centering
    \includegraphics[width=0.42\textwidth]{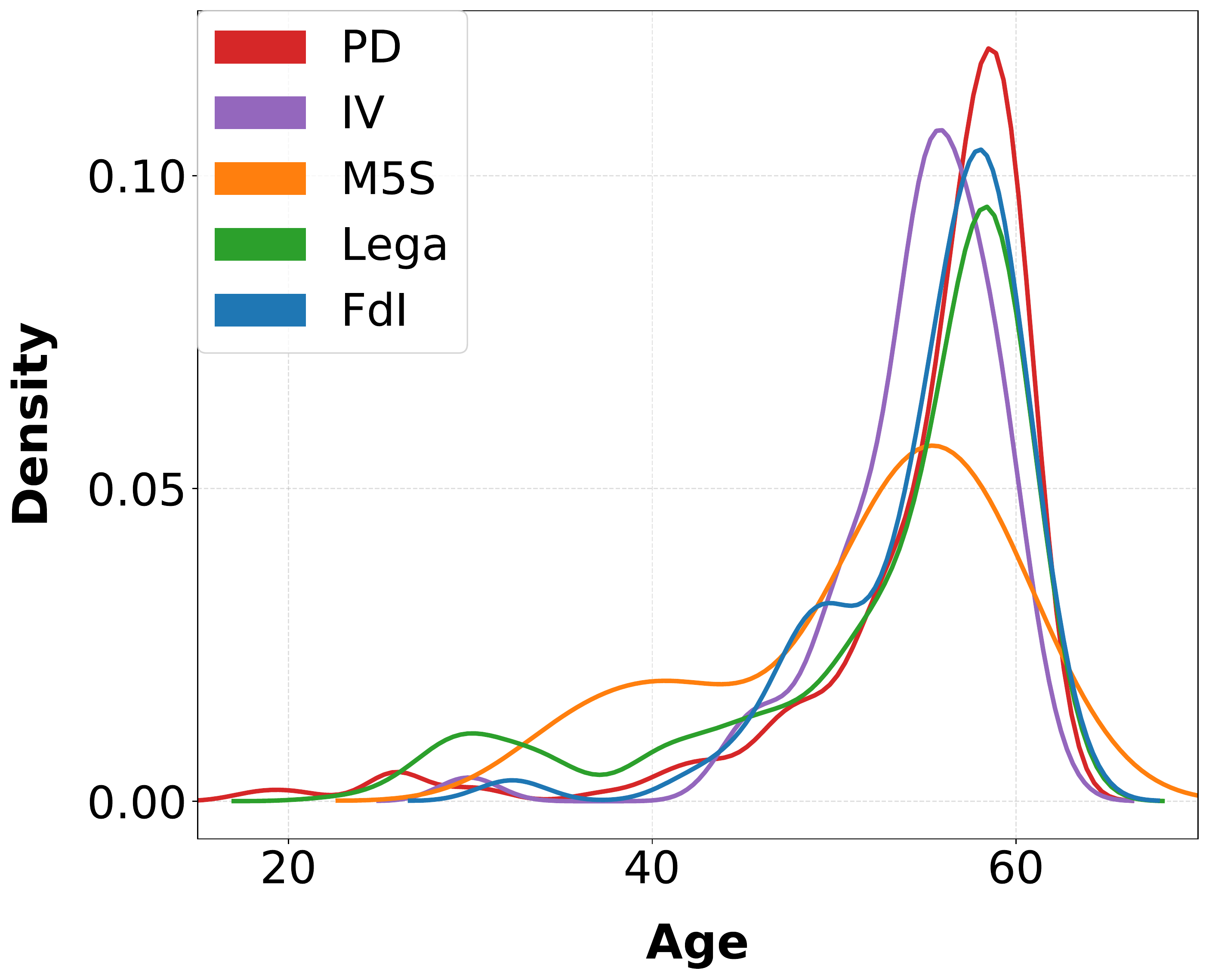}
    \hspace{1cm}%\qquad
    \includegraphics[width=0.42\textwidth]{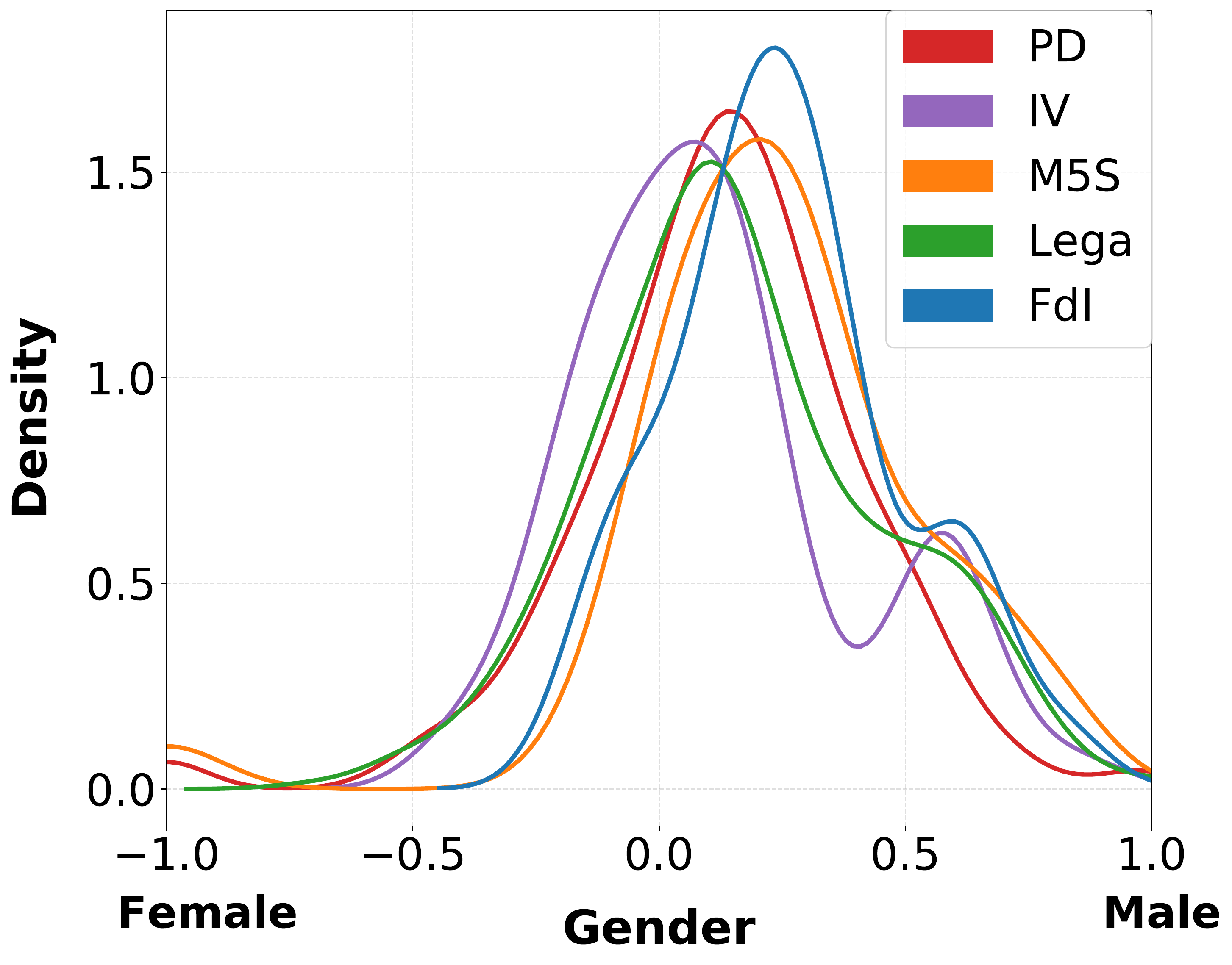}
    \caption{Probability density functions of age (left) and gender (right) impressions by party computed by kernel density estimation.}
    \label{fig:kde_gender_distribution}
\end{figure}

In this section, we ask whether the parties tailor their message around migration to specific audiences in terms of age, gender, and geography.
Note that the data we collect tells us the final reach of the ads, we cannot distinguish between explicit targeting of the authors and the platform optimization of the ad delivery to potentially interested audience.
Nevertheless, this gives us a precise measure of the final reach of the ads, and as such it is still a strong clue for investigating the ads' target audience.
Figure \ref{fig:fireworks} (left) plots a circle for each ad, located at the mean age and gender of audience reached, colored by the authoring party, and sized by impressions.
For each party, we also plot the global impression-weighted average as an ellipse, such that each axis reflects one standard deviation of age and gender.
Overall, we find most ads to be targeting older and more male audiences.
However, there are differences among the parties, which we compare in Figure \ref{fig:kde_gender_distribution} by plotting kernel density estimates of impressions for each party over age (left) and gender (right).
We find M5S to have a broader age distribution, having more young audience, followed by IV.
Considering gender, IV captures more female audience, while FdI is more male-oriented.

\begin{figure}[t]
    \centering
    \includegraphics[width=0.46\textwidth]{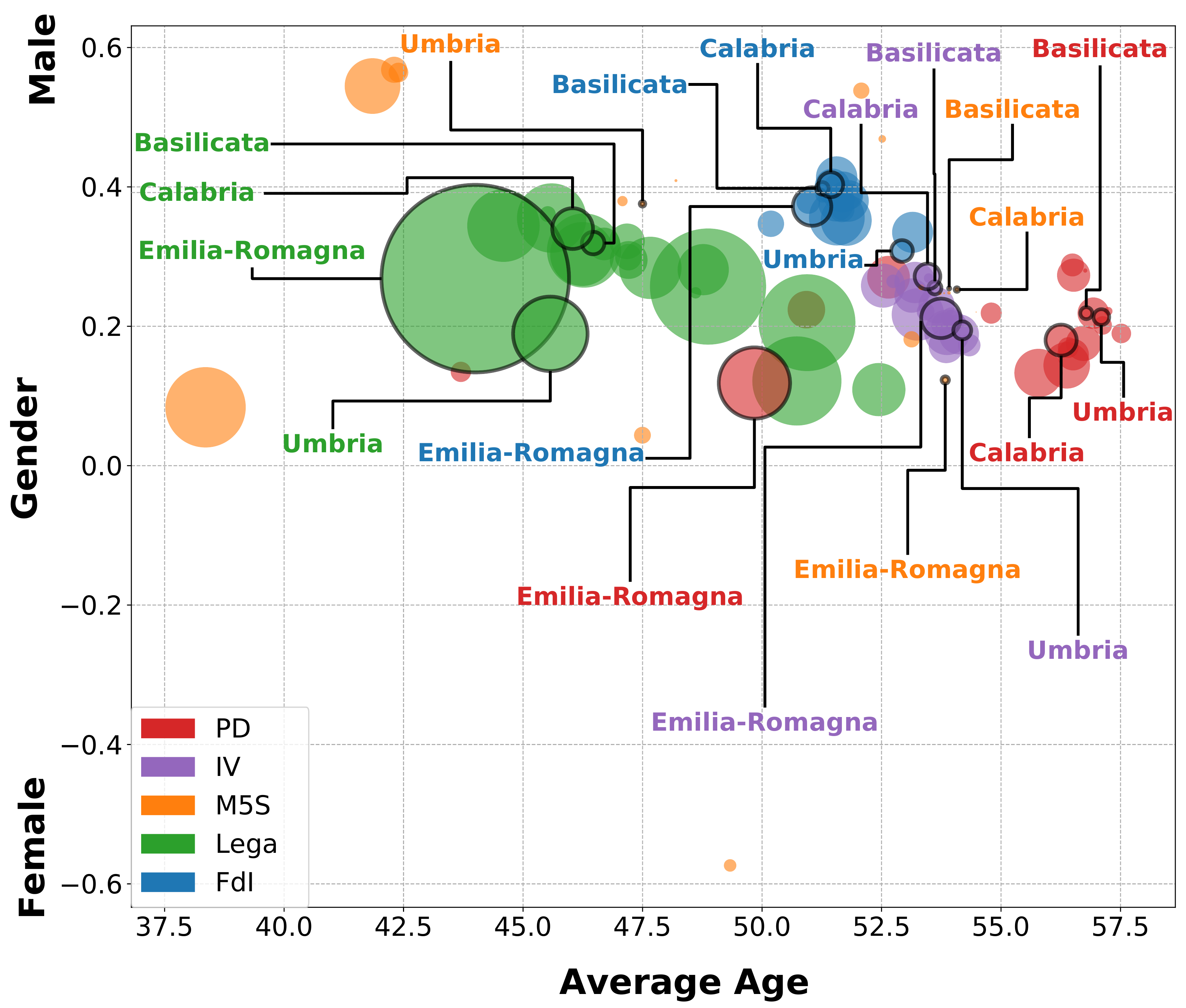}\hspace{0.5cm}
    \includegraphics[width=0.46\textwidth]{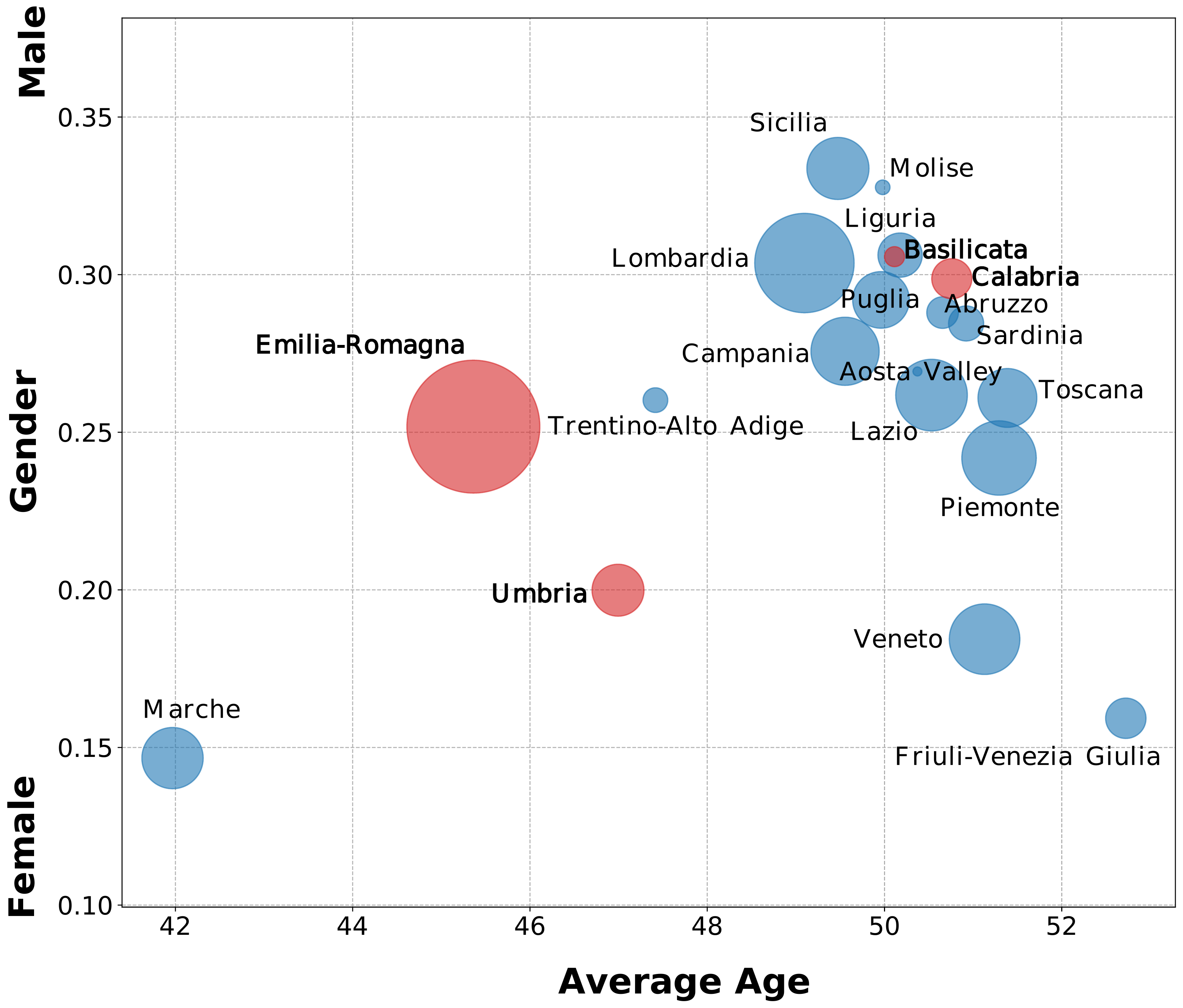}
    \caption{Average reach of ads by region in terms of age and gender, size proportional to impressions. On the left we group ads by region and party, while on the right we group them only by region (in red are regions with regional elections).} \label{fig:scatter_per_region}
    \vspace{-\baselineskip}
\end{figure}

% Sum of all authors relative to Lega:
    % spend = 87314
    % impressions = 13807805

% Sum of all authors relative to PD:
    % spend = 13786
    % impressions = 3022386

% Sum of all authors relative to FdI:
    % spend = 10963
    % impressions = 2695468

% Sum of all authors relative to M5S:
    % spend = 3533
    % impressions = 1532483

% Sum of all authors relative to Italia Viva:
    % spend = 6023
    % impressions = 2217973

% RATIO Euros per impression
% PARTY

% MIN. min:  ('Democratic Party', 0.00037934720037024284)
% MIN. max:  ('Lega', 0.002229075218796934)

% AVG. min:  ('Five Star Movement', 0.00230573444999571)
% AVG. max:  ('Lega', 0.006323566175942402)

% MAX. min:  ('Five Star Movement', 0.006295064377682403)
% MAX. max:  ('Democratic Party', 0.014215654077723043)

Considering the geography, in Figure \ref{fig:scatter_per_region} (left) we aggregate the impressions of ads by different parties by the region, such that one ad may contribute to several regions.
In particular, each circle is a regional reach of a party, scaled by the number of impressions and located at the average age and gender of the targets.
Note the plot is zoomed in, such that the axes span smaller interval than \ref{fig:fireworks}.
We find that the points cluster by party (color), implying that the targeting is mostly differentiated by party, instead of geography.
This finding suggests that the targeting strategy originates from the party focus and it is only slightly adjusted for each region.
In fact, plotting the aggregated target audiences by region in Figure \ref{fig:scatter_per_region} (right), we find that most of the difference among regions are due to events such as elections.
In fact, elections seem to make the targeting more gender-balanced and younger in age.

\begin{figure}[t]
    \centering
    \includegraphics[width=\textwidth]{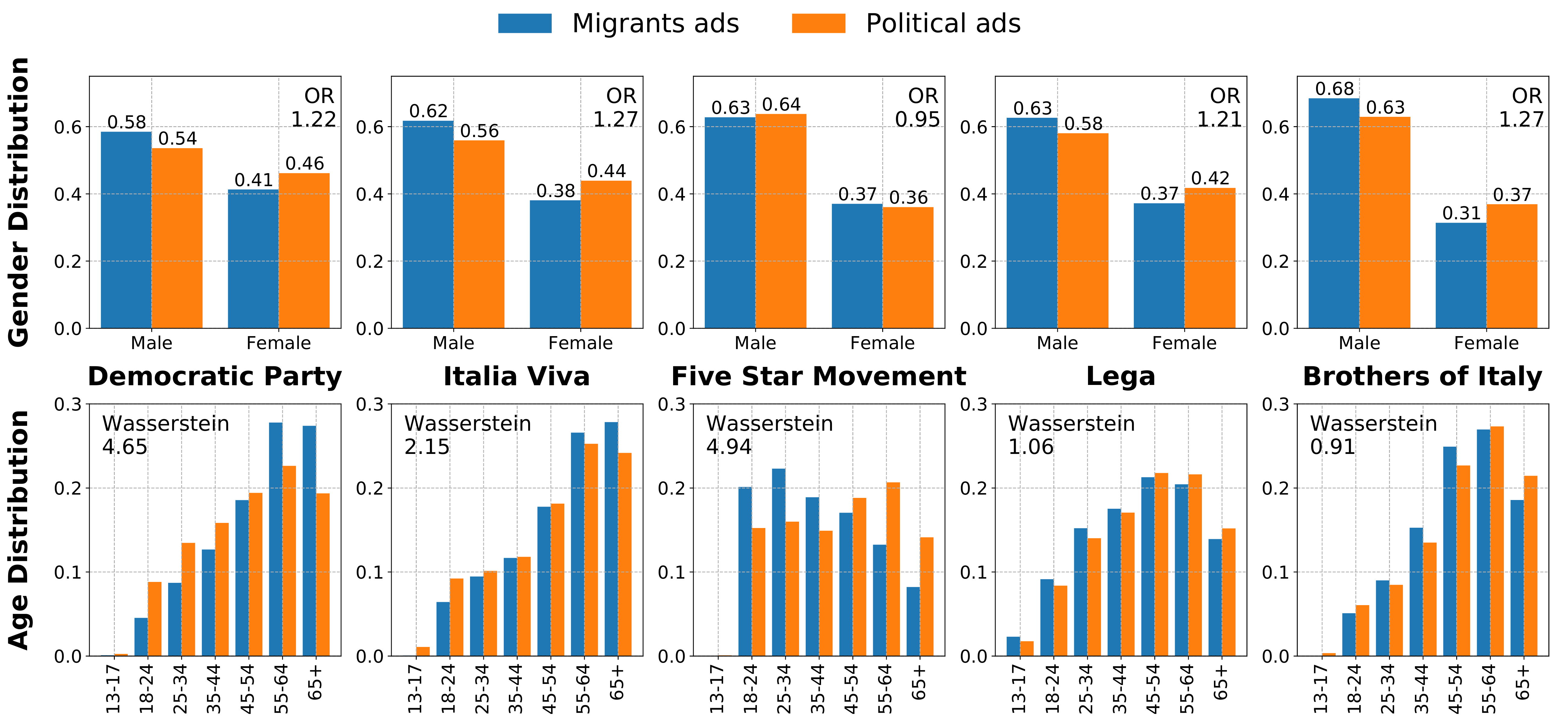}
    \caption{Distributions of impressions for migrant-related and all political ads over gender (top) and age (bottom), per party.}
    \label{fig:prob_age_gender_baseline}
    \vspace{-\baselineskip}
\end{figure}

Next, we ask whether this targeting is due to the specific topic of immigration, as compared to overall political ads.
Recall that we collect all other ads posted by the politicians present in our collection, making up a topic-neutral baseline (note that our view of political parties is thus constrained by the authors we encounter in immigration-specific collection).
Figure \ref{fig:prob_age_gender_baseline} shows the distribution of impressions for migrant-related and all political ads over gender and age, per party.
In general, we find that all parties are reaching different sub-populations when advertising about migration, either by gender or age.
Focusing on gender, we compute the odds ratio of the two binomial distributions.
All parties, except M5S, have higher odds of targeting males on migrant-related ads (significant at $p<0.0001$ using Fisher's exact test), compared to the general political ads, on average about 20\% higher.
For age, Fisher's exact test shows the distributions to be significantly different at $p<0.0001$.
Instead to assess the effect size, we compute Wasserstein distance (also known as ``earth mover's distance'') between the two age distributions.
The higher this metric, the more one needs to ``move'' from one distribution to achieve the other.
We find M5S and PD to have the most different migrant-related ad audience, compared to that of all of their political ads, with the focus shifting drastically from older to younger segments for M5S.

%We find that overall there is not much difference in the audience reached by each party when focusing on the topic of immigration.
%However, we can observe differences for some parties.
%For instance, immigration-related ads have a more male-oriented audience in the case of the far right party Brothers of Italy (\inote{differences of medians}); it is instead more gender balanced than baseline ads in the case of the centrist party Italia Viva.
%We can also note that for Lega the median age of immigration-related ads is higher than the baseline.
%These findings are consistent with survey-based results that suggest a larger concern about immigration among older males (\inote{TODO: citation!}).
%\mynote{New plots: gen\_dist and age\_dist. Now for example Italia Viva is not more gender balanced than others.}

% MEDIAN
% BROTHERS OF ITALY
% migrants:
% 0.623578
% baseline:
% 0.586773

% LEGA
% migrants:
% 0.563264
% baseline:
% 0.552764

% FIVE STAR MOVEMENT
% migrants:
% 0.609146
% baseline:
% 0.647721

% DEMOCRATIC PARTY
% migrants:
% 0.566722
% baseline:
% 0.567033

% ITALIA VIVA
% migrants:
% 0.5244139999999999
% baseline:
% 0.5751935

%\inote{For each Ad I've a gender value between -1 and 1 and an age value between 15 and 70. First I've rounded the gender value of each Ad to one number after comma (decimal place) and the age value to zero numbers after comma. Then I've found that maybe a density plot using kernel density as estimation is a better solution.}

% Was: Probability density function of age (above) and gender (below) impressions, for immigration-related ads and all political ads for each party.

%Party-specific demographic targeting. Compare number of impressions per party and number of votes received by the party, can do this for migrants subset or general ads (good if we can get per-region). If the ad targeting coincides with voters, they are ``preaching to their own choir", and if not, they are trying to reach new voters.

% https://www.termometropolitico.it/1434282_elezioni-europee-sesso-eta.html

%\mynote{Why are some costs per impression so different than others? Could it be because they are micro-targeting and this drives up the price? Measuring micro-targeting: 2 entropies: Entropy of impressions over demographic buckets and over geographic buckets. Then we correlate these measures to Euros per impression. (could also combine using harmonic average, but for that each of these entropies need to be normalized by log n (n being number of buckets))}

Lastly, we wonder whether a higher cost per impression may be due to narrower targeting of the viewership.
To test this hypothesis, we measure the extent of micro-targeting by computing entropy of the distribution of impressions over demographic buckets and over geographic regions.
Correlating this measure with cost per impression, we find the correlation between cost per impression and demographics entropy to be $\rho_d=-0.112$, and correlation between cost per impression and geographic entropy  $\rho_g=-0.243$, which indicates very little relationship between the narrowness of the targeting and efficiency.
%%%%%%%% HOW SHOULD WE FRAME THIS DIFFERENTLY?

\section{Discussion}

In this work, we have analyzed the behavior of political and social advertisers on Facebook on the topic of immigration in Italy.
% We focused our analysis on a one year period, from March 2019 to March 2020, and queried the Facebook Ad Library API% with a set of immigration-related keywords
% , obtaining a set of \num{2312} ads.
% We used WikiData to analyse the authors of these ads: we were able to find and validate which of these ads belonged to politicians, and to which party they adhere.
% For the political authors, we also gathered all their ads in the same period, obtaining a baseline set of \num{17 014} political ads.
We believe our data set, including the ad metadata and enrichment by using external sources, to be of value to the community, and thus we make it openly available%, as per Facebook Ads Library Terms of Service
.\footnote{\url{https://github.com/PotenteOpossum/Facebook-Ads-Politics-of-Migration-in-Italy}}

In our analysis, we find evidence that Facebook advertising is considered important by many political actors.
Among the top four authors by total number of impressions, we find that three are party leaders.
In particular, Matteo Salvini (leader of the Lega party) is the top one, both by impressions and by total expenditure.
This results reflects two main characteristics of his leadership: a focus on nativist discourse~\cite{bulli2018immigration} and the usage of social media as a political platform.

% We analyzed the temporal and geographical distribution of ads impressions.
We find ads impressions to be largely correlated with national and local elections, thus corroborating the hypothesis that they are viewed as an important means to attract votes.
In particular, we find the largest peak to coincide with European elections.
The possibility of running advertising campaigns in the last days before elections has been put under scrutiny in all Europe~\cite{council2018internet}, where some countries (including Italy) have rules that impose electoral silence on other media -- rules which do not apply to this increasingly important medium.
Indeed, we observe that geographically the largest number of impressions per capita is in the region of Emilia-Romagna, which hosted a regional election on 26th January 2020.
This election has been recognized by political commenters as having a wider national importance~\cite{emiliaromagna}.
We also find that immigration-related ads are most prevalent in the North of Italy, as well as in Sicily and Sardinia.
While the former could be connected to those regions being strongholds of Lega, the latter result is more surprising and could be related to overestimation of the perceived presence of immigrants \cite{council2018internet}.

We then turned our attention to the controversial phenomenon of micro-targeting.
The political ads we observe tend to target older and more male users.
We observe significative differences across parties: for instance, M5S targets a younger audience, while FdI a more male one.
This is consistent with the demographic distribution of their voters, as estimated from surveys \cite{formigoni2018elezioni}.
% We analyze how such micro-targeting changes w.r.t. to geographical distribution;
However, we find that regional elections can shift the audience towards the population average, both in terms of age and gender.
Furthermore, the ads targets change with the topic: some parties appear to reach a significantly different audience when speaking about immigration.
In particular, most parties tend to reach a more male-oriented audience when speaking about immigration, while in terms of age each party has a different bias compared to their general audience.
We believe these results to be an indication of micro-targeting, i.e., parties trying to reach different sub-populations with different messages~\cite{ribeiro2019microtargeting}.
% Finally, we characterized each party's main messages by comparing the main words they used in their ads.

\paragraph{Limitations.}

We restricted our attention to a specific use case of political advertising on Facebook: the topic of immigration in Italy.
Since immigration has become one of the most loaded issues in Italian party politics~\cite{bulli2018immigration}, our findings are limited to this specific context.
Narrowing our scope in topic, time, and space, however, has the beneficial effect of removing potential confounding factors.

Some specific aspects of the advertising campaigns are not available to the public through the Facebook Ads Library API, thus limiting our findings.
In particular, different mechanisms could be driving the micro-targeting we observed:
on the one hand, a certain sub-population could be engaging more with immigration-related ads, thus increasing their impressions;
on the other, they could be explicitly targeted by the parties.
For instance, Hegelich et al.~\cite{hegelich2019microtargeting} find that ``organic'' interactions -- those resulting from free posts that are usually disseminated to existing party followers -- may attract more views than paid advertising.
Without further data shared by the platform, there cannot be a conclusive answer to this important question.
It has also been reported that the API appeared unstable and unreliable \cite{mozilla2019data}. To test the stability of our data, we re-query both the migrant-related ads and general political ones two weeks after, and find that 1 ad is missing in the former collection and 3 in latter. Thus we conclude that, in our case, the API is fairly stable.

Despite these limitations, we believe our work brings important elements in the debate on Facebook advertising in politics, showing its importance in electoral campaigns, and bringing statistical significance to the hypothesis that different parts of the population are being targeted by different ads.
More work is necessary on this topic.
In order to understand the different political messages, it would be possible to use text and image mining to characterize ads, and understand which views and behaviors they promote.
Furthermore, comparing the data we obtained from Facebook with other data sources -- e.g., surveys, news coverage, election results -- could help in understanding the complex relationship of these ads with real-world events.

%\begin{table}
%\caption{Table captions should be placed above the tables.}\label{tab1}
%\begin{tabular}{|l|l|l|}
%\hline
%Heading level &  Example & Font size and style\\
%\hline
%Title (centered) &  {\Large\bfseries Lecture Notes} & 14 point, bold\\
%1st-level heading &  {\large\bfseries 1 Introduction} & 12 point, bold\\
%2nd-level heading & {\bfseries 2.1 Printing Area} & 10 point, bold\\
%3rd-level heading & {\bfseries Run-in Heading in Bold.} Text follows & 10 point, bold\\
%4th-level heading & {\itshape Lowest Level Heading.} Text follows & 10 point, italic\\
%\hline
%\end{tabular}
%\end{table}

%\begin{figure}
%\includegraphics[width=\textwidth]{fig1.eps}
%\caption{A figure caption is always placed below the illustration.
%Please note that short captions are centered, while long ones are
%justified by the macro package automatically.} \label{fig1}
%\end{figure}

%\bibliographystyle{splncs04}

\small
\bibliographystyle{splncs04nat}
\bibliography{migrationads}

\clearpage
\appendix
\section{Appendix}
\label{sec:appendix}

\begin{table}[h]
\centering
\caption{Top 10 words for each party by odds ratio of presence in migrant ads compared to all political ads (translated to English).}
\label{tab:top_words_per_party}
\small %\scriptsize
\begin{tabular}{p{2.1cm}p{2.1cm}p{2.1cm}p{2.1cm}p{2.1cm}}
 \toprule
PD & IV & M5S & Lega & FdI\\
 \midrule
PD & hate & movement & Lega & Europe \\
Salvini & Italians & Italy & Salvini & Italy \\
politics & September & countries & clandestines & brothers \\
migrants & Matteo & Conte & Italians & law \\
Italy & death & answer & truth & clandestines \\
party & voice & policies & migrants & territory \\
Europe & Mario & Lega & Matteo & NGO \\
rights & Salvini & Europe & immigrants & activity \\
law & community & agreement & immigration & nations \\
women & security & migrants & ports & citizenship \\

% pd & odio & movimento & lega & europa \\
% salvini & italiani & italia & salvini & italia \\
% politica & settembre & paesi & clandestini & fratelli \\
% migranti & matteo & conte & italiani & legge \\
% italia & morte & risposta & verita & clandestini \\
% partito & voce & politiche & migranti & territorio \\
% europa & mario & lega & matteo & ong \\
% diritti & salvini & europa & immigrati & attivita \\
% legge & comunita & accordo & immigrazione & nazioni \\
% donne & sicurezza & migranti & porti & cittadinanza \\
 \bottomrule

\end{tabular}
\end{table}

\paragraph{List of ad authors manually removed:}
% `Xương Khớp Ông Bảo',
`Move To Canada Today',
`Patagonia',
`VisaPlace - Niren \& Associates Immigration Law Firm',
`Immigration Spot',
`Battlefield Italia-La pagina',
`Videodrome'.

\paragraph{Keywords used to retrieve ads from the Facebook Ads Library:}
`migrante', `migranti', `immigrato', `immigrati', `immigrata', `immigrate', `ius soli', `ius culturae', `sbarchi', `sbarco', `migrazioni', `migrazione', `clandestino', `clandestini', `clandestina', `clandestine', `profugo', `profughi', `profughe', `profuga', `scafisti', `scafista', `extracomunitario', `extracomunitari', `extracomunitaria', `extracomunitarie'.

\end{document}